\documentstyle[11pt,paspconf,epsfig]{article}
 
\begin{document}

\title{A Test of Tully-Fisher Distance Estimates Using Cepheids and Type Ia Supernovae.}

\author{T. Shanks} \affil{Dept. of Physics, University of Durham, South Road, Durham
DH1 3LE, England}

\begin{abstract} We update and extend the results of Shanks (1997) by making a direct
test of Tully-Fisher distance estimates  to thirteen spiral galaxies with  HST
Cepheid distances and to ten spiral galaxies with Type Ia supernova (SNIa) distances.
The results show that the Tully-Fisher distance moduli are too short with respect to
the Cepheid distances  by 0.46$\pm$0.11mag and too short  with respect to the SNIa
distances by 0.49$\pm$0.18mag. Combining the HST Cepheid and the best SNIa data suggests that,
overall, previous Tully-Fisher distances at v$\sim$1000 kms$^{-1}$ were too short by
0.43$\pm$0.09mag, a result which is significant at the 4.6$\sigma$ level. These data
therefore indicate that previous Tully-Fisher distances should be revised upwards by
22$\pm$5\% implying, for example,  a Virgo distance  of 19.0$\pm$1.8Mpc. The value of
H$_0$ from Tully-Fisher estimates is correspondingly revised downwards from
H$_0$=84$\pm$10kms$^{-1}$Mpc$^{-1}$ to H$_0$=69$\pm$8kms$^{-1}$Mpc$^{-1}$. There is
evidence that the Tully-Fisher relation at large distances is affected by Malmquist
bias. In this case, we argue that H$_{0}<$50kms$^{-1}$Mpc$^{-1}$ cannot be ruled out 
by Tully-Fisher considerations.

\end{abstract}


\section{The Importance of Measuring Hubble's Constant.}

Since part of our motivation for investigating H$_0$ arises from its implications for
$\Omega_0$, we begin by repeating the reasons for the theoretical preference for a
flat or (just) closed Universe. In the absence of a cosmological constant, this means
that the Universe has high mass density and the first argument for such an $\Omega$=1
model was given by Einstein \& de Sitter (1932) who argued that given the Universe is
spatially flat locally, in the absence of further information we might expect that
the Universe is also spatially flat globally. Following Einstein (1918), Wheeler
(1964) and  Lynden-Bell et al (1995) have further suggested that Mach's Principle can be
more easily accommodated in General Relativity in a (just) closed model rather than in
a spatially infinite model because, in the words of Lynden-Bell et al, `If we add to
Einstein's theory the requirement that space must be closed....the conundrum of
inertia in an almost flat, almost empty space does not occur!'  Another problem with
open models was pointed out by Ya.B. Zel'dovich who suggested that if the
Universe appeared from a quantum fluctuation then this might be more plausible in a
spatially finite Universe rather than a spatially infinite Universe. The more recent
arguments of quantum cosmology also appear to leave  the choice of  a model where
$\Omega_0$=1 exactly or an unphysical model where $\Omega_0$=0 exactly (Turok \& Hawking,
1998).

This  context of theoretical preference for spatially flat models was reinforced by the
success of the inflation model (Guth, 1981, Linde, 1982) in solving  the `flatness
problem' at the same time as the `horizon problem'.  The `flatness problem' harks back
to the paper of Einstein \&  de Sitter; to observe an $\Omega_0$ which is lower than
unity at the present day, it is required to postulate that $\Omega$ in the early
Universe was microscopically different from unity. To avoid the  fine-tuning of the
density parameter that this requires, inflation suggests that there was a period of
exponential expansion at early times which reduced any initial spatial curvature to
essentially zero, leaving $\Omega_0$=1 to high accuracy at the present day. This
argument partly motivated the introduction of CDM (Peebles, 1982, Blumenthal
et al, 1982) because standard baryon nucleosynthesis arguments seem to suggest that the
baryon density has to be low.

The possibility  that we may live in a low density Universe with a cosmological
constant, which again would give  us the above theoretical advantages of living in a
spatially flat or just closed Universe, has been considered by Peebles (1984)
and others. However, as Peebles also notes, this argument appears circular with a high
degree of  fine tuning required in the early Universe to arrange for inflation to
clean out all the vacuum energy into radiation but to leave one part in 10$^{120}$ to
allow us to observe a cosmological constant at the present day. Also the clear
view of particle physicists is  that they would prefer the cosmological
constant to be either so huge as to be cosmologically impossible or zero (eg Kolb,
1998).

On the basis of the above arguments, the current standard cosmological models are
therefore complicated, finely tuned affairs - either open or  (just) closed with
a  small cosmological constant. Invoking CDM in either case introduces further
complication. As noted by Peebles (1984), the problem is that the baryon and CDM
densities then lie within 1-2 orders of magnitudes of each other which again implies
fine-tuning  given there is no physical reason for this coincidence. Even those
sceptical of fine-tuning arguments might still regard it as circular reasoning to
start with inflation to explain a fine-tuning coincidence only to have to introduce
CDM and another fine-tuning coincidence at a later stage. Put another way, the
introduction of CDM helps solve the overall flatness problem but leaves us looking at
the baryon flatness problem which is why is $\Omega_{baryon}$ to within 1-2 orders of
magnitude of unity?

Shanks (1985) and Shanks et al (1991) therefore suggested that the simplest model
would have $\Omega_{baryon}$=1. In this case there is neither fine-tuning of the
overall density parameter, the cosmological constant nor the baryon and CDM matter
densities. However, this means that something else has to give and the view that was
taken by Shanks (1985) was that it might be Hubble's Constant that  was the problem.
After all, distance scale estimates of H$_0$ have been highly uncertain, with
Hubble's final value in 1953 being H$_0\sim$ 500kms$^{-1}$Mpc$^{-1}$.

Three advantages immediately  accrue if the value of Hubble's Constant turns out
to be below today's usually quoted 50-100 kms$^{-1}$Mpc$^{-1}$ range. First,
Shanks (1985) quoted  0.01$<\Omega_{baryon}h^2<$0.06 as the allowed range for
H$_0$ from nucleosynthesis considerations. Now the upper limit here has
oscillated over the years as the observed abundances of Helium-4  and Deuterium
have changed (Izotov et al, 1997, Burles \& Tytler, 1998). However, the above
range may still do no disservice to the real observational uncertainties when
systematics are taken into account and therefore  if
H$_0\sim$25kms$^{-1}$Mpc$^{-1}$ then a model with $\Omega_{baryon}$=1 would be
compatible with the nucleosynthesis constraints. Second, Shanks (1985) noted that
there is a significant mass of X-ray gas in the Coma cluster; this culminates in
the problem known as the `baryon catastrophe' in the Coma cluster where CDM
models with $\Omega_0$=1 cannot explain Coma's high baryon fraction (White et al,
1993). Shanks (1985) pointed out  that the mass of the gas had a much stronger
dependence on H$_0$ than the virial mass, leading to the ratio
M$_{Virial}$/M$_{X-ray}$ = 15h$^{1.5}$. So whereas the X-ray gas is a factor of
fifteen away from explaining the virial mass for H$_0$=100kms$^{-1}$Mpc$^{-1}$ and a
factor of five for H$_0$=50kms$^{-1}$Mpc$^{-1}$, if H$_0$=25kms$^{-1}$Mpc$^{-1}$
then the X-ray gas is within a factor of two of the virial mass of Coma. Finally,
any $\Omega_0$=1 model would prefer a low value of H$_0$, given that the age of
the Universe at 6.5-13 Gyr is uncomfortably close to the ages of the oldest stars
if H$_0$ lies in its usual 50-100kms$^{-1}$Mpc$^{-1}$ range. If a CDM model with
$\Omega_0$=1 demands a low H$_0$ anyway on timescale grounds, then the view might
be taken that the above arguments could make the introduction of CDM redundant.
Therefore if the Hubble Constant was lower  than 50kms$^{-1}$Mpc$^{-1}$ then
there would be considerable advantage for an $\Omega_{baryon}$=1 model, with no
finely tuned $\Omega_{CDM}:\Omega_{baryon}:\Omega_{\Lambda}$ ratios to explain,
motivating renewed investigations to see whether current cosmological distance
scale estimates of H$_0$ are correct.

\section{Testing the Tully-Fisher Estimate of H$_0$.}

For many years the Tully-Fisher (TF) relation (Tully \& Fisher, 1977) has been a
principal argument for a high Hubble's constant. This argument was strong  because
the TF relation was calibrated using ground-based Cepheid distances to Local Group
spirals and so used less distance scale steps than other methods. However, although
there have been criticisms of the TF method (Sandage 1994 and refs. therein), a
definitive test of the accuracy of the TF distances required Cepheids to be
discovered in more distant galaxies.

Here,  we use thirteen newly available Cepheid distances  from the HST Distance Scale
Key project (Freedman, 1997, Turner et al,1998, and refs. therein) and from other HST
(Tanvir et al, 1995, Sandage et al, 1996 and refs. therein) observations to test TF
distance estimates for more distant galaxies than has previously been possible.  We
shall supplement the Cepheid-TF data with ten SNIa-TF galaxies, ie galaxies which
have both SNIa distances and  TF distances, which  provides a further
way to test the calibration of the TF scale.  

\begin{figure} 
\centerline{\epsfig{figure=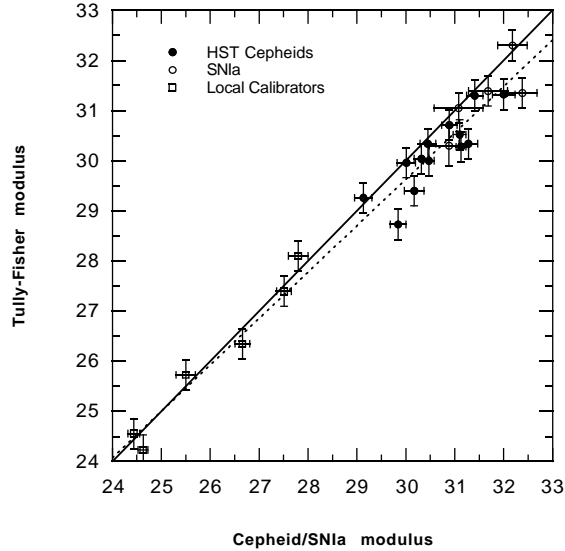,height=3in}}
\caption{The comparison of Tully-Fisher distance moduli with HST Cepheid/SNIa(`best'
sample) moduli and ground-based Cepheid moduli at lower distances. The solid
line shows the 1-1 relation and the dashed line is the least squares fit. The
Tully-Fisher distances are 22$\pm$5\% too short comparied to the HST Cepheid/ SNIa
distances.} 
\end{figure}

Table 1 of Shanks (1997) lists  eight spirals for which Cepheid distances  have been
obtained and with inclination, i$>$40 deg, which means they have good TF distances. 
We note that the TF distance modulus for NGC1365 in Table 1 is in error and should
read 30.34$\pm$0.3. We have also included 5 further galaxies NGC3627, NGC2090,
NGC2541, NGC7331, NGC4414 whose Cepheid distances have been published by the Key
Project team more recently. (Turner et al, 1998 and refs. therein).

Table  2 of Shanks (1997) shows a further twelve galaxies which have both TF and SNIa
distances. Where galaxies have both an SNIa and a Cepheid distance, the default has
been to include them in the Cepheid-TF list of Table 1. NGC3627 and NGC4414 now have 
Cepheid distances and have therefore been included with the galaxies of Table 1.
SN1937D, SN1957A and SN1976B have large dust extinctions. SN1971I and SN1991T are
doubtful Type Ia on the basis of their spectra (D. Branch, priv. comm.) which leaves
a `best' sample of five SNIa. The SNIa distances have been calibrated using five galaxies
with SNIa and Cepheid distances  which gives M$_B$(max)=-19.38$\pm$0.11 (Shanks,
1997).  Most of the SNIa in Table 2 of Shanks (1997) do not have accurate enough
light curves to allow use of the proposed maximum luminosity-decay rate correlation
(Hamuy et al, 1996). The Tully-Fisher distances mostly come from Pierce (1994)
and otherwise they are calculated following the precepts of Tully \& Fouque (1985).

Fig. 1 shows the comparison between Tully-Fisher distances and HST-based
Cepheid/SNIa distances (`best sample'). Also shown are the  six galaxies with ground
based Cepheid distances which formed the previous Local Calibrators for the TF
relation. We see that there is a significant difference between the TF and
Cepheid/SNIa distances for galaxies with (m-M)$>$29.5. Taking the thirteen  galaxies
with HST Cepheid distances  and with i$>$40deg, we find that the TF-Cepheid
difference corresponds to  $\Delta$(m-M)=0.46$\pm$0.11 or a 4.2$\sigma$ discrepancy.
Similarly taking the full sample of ten TF galaxies with SNIa we find
$\Delta$(m-M)=0.49$\pm$0.18 or a 2.7$\sigma$ discrepancy. Finally, taking the
thirteen HST Cepheid galaxies and the five SNIa galaxies in the `best' sample we obtain
an overall discrepancy of $\Delta$(m-M)=0.43$\pm$0.09 or a 4.6$\sigma$ discrepancy. The
further inclusion of the original 6 calibrators reduces the discrepancy to 
$\Delta$(m-M)=0.33$\pm$0.08 which is still a  4.0$\sigma$ discrepancy based now on 24 galaxies.

Thus TF distances at v$\sim$1000 kms$^{-1}$ are systematically too short by
22$\pm$ 5\%. The TF distance to Virgo of Pierce \& Tully (1992)  therefore
requires a correction from 15.6 Mpc to 19.0 $\pm$1.8 Mpc. Assuming their Virgo
infall model, the Pierce \& Tully value of H$_0$ at Virgo is therefore decreased
from H$_0$=84 to H$_0$=69$\pm$8 kms$^{-1}$Mpc$^{-1}$. Thus far there is little
controversy in this result, with Giovanelli et al (1997) and Hughes et al (1998) 
for the HST Key Project Team coming to approximately the same result for H$_0$.
The difference between the old calibrators and the new can be clearly seen in the
TF relation shown in Fig. 10 of Hughes et al (1998). This correction to the value
for H$_0$ is in the right direction for $\Omega_{baryon}$=1 but it is clearly not
yet enough.

\begin{figure}
\centerline{\epsfig{figure=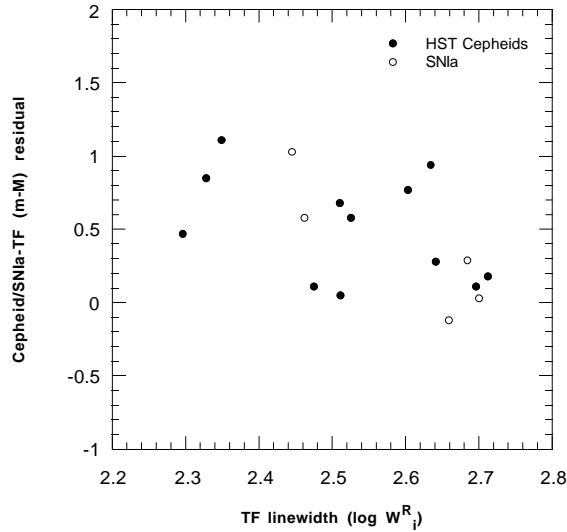, height=3in}}
\caption{The correlation of Cepheid/SNIa-TF distance moduli residuals with TF
linewidth for distant galaxies with (m-M)$>$29.5. The data tentatively indicate 
at the 2$\sigma$ level that the lowest linewidth galaxies show the biggest
residuals and this is a possible signature that TF distance estimates are affected
by Malmquist bias.} 
\end{figure}

However, it is interesting to consider the possible reasons for the change in the
zeropoint of the TF relation. Hughes et al consider that the only problem may
have been the  poor statistical precision  of the  previous calibration which was
only based on six galaxies. However, it has been claimed previously that the TF
relation may be affected by Malmquist bias (Sandage 1994 and references therein)
and it has to be said that Fig. 1 does give the impression of there being a
scale-error in the TF relation rather than simply a zeropoint error. The least
squares fitted line in Fig. 1  gives a 2.1$\sigma$ rejection of a null hypothesis
of  a slope of unity; this is evidence that the TF error is distance dependent.
Sandage (1994) has further predicted that a signature of Malmquist  bias in the
TF relation would be if the value of H$_0$ was a function of the line-width of
the galaxy. In Fig. 2, which shows the correlation between Cepheid/SNIa-TF
residuals and TF linewidths for galaxies with (m-M)$>$29.5, we find that there is
some evidence for this effect with low linewidth galaxies showing the biggest
($\sim$0.7mag) residuals. Again the effect is at the 2$\sigma$ level and is to be
confirmed in larger samples. Of course, if the Cepheid/SNIa-TF error is caused by
Malmquist bias then the 22$\pm$5\% effect noted here may be a lower limit to the
actual error in the TF scale at larger distances. Since the TF relation has been
used as far as  the Coma cluster and beyond, the further validation of the TF
distance scale may await the detection of Cepheids in spirals at the distance of
the Coma cluster; such observations would be  possible from a Next Generation
Space Telescope with diffraction limited performance in the visible wavebands.

\section {Conclusions \& Discussion.}

Our conclusions are as follows: 

\begin{itemize}

\item Cepheid and SNIa observations suggest that Tully-Fisher distances at v$\sim$
1000kms$^{-1}$ are too short by 22$\pm$5\%.

\item The  Pierce \& Tully (1992) Tully-Fisher distance to Virgo  therefore increases
from 15.6 Mpc to 19.0$\pm$1.8 Mpc.

\item The corresponding Pierce \& Tully estimate of H$_0$ at Virgo therefore
decreases from H$_0$=84 to H$_0$=69$\pm$8 kms$^{-1}$Mpc$^{-1}$, assuming their 
Virgo infall model.

\item Tully-Fisher distances may be affected by Malmquist bias which implies that the
TF-Cepheid/SNIa discrepancy could worsen at larger distances.
 
\end{itemize}

We finally consider whether the real value of H$_0$ could still be significantly lower
than  that quoted above. If we take the view that the TF relation at large distances
may be affected by Malmquist bias then we may only trust TF estimates of H$_0$ at
v$\sim$1000 kms$^{-1}$. Then we take our revised TF Virgo distance of 19.0$\pm$1.8Mpc
and apply the 17\% Cepheid zeropoint correction of Feast \& Catchpole (1997) based on
Hipparcos parallaxes of Cepheids and the likely effects of metallicity. This then
results in a Virgo distance of 22.2$\pm$2.5 Mpc. Assuming neither our infall velocity
into Virgo nor the peculiar motion of Virgo we naively take the heliocentric velocity
of Virgo of 1016$\pm$42 kms$^{-1}$ of Pierce \& Tully (1992) which results in an
estimate of H$_0$=46$\pm$5kms$^{-1}$Mpc$^{-1}$. Now clearly the error here does not
take into account the large systematic error caused by our ignorance of  the peculiar
velocities. And certainly if we repeat the above experiment with the revised TF
estimates of the distance to Fornax we then obtain H$_0$=65$\pm$7kms$^{-1}$Mpc$^{-1}$;
with Ursa Major we obtain H$_0$=43$\pm$5kms$^{-1}$Mpc$^{-1}$. Clearly the effects of
peculiar velocity  are important and are causing significant errors in our H$_0$
estimate. However, our only aim here is to point out that the range of H$_0$ estimates
from the Hipparcos/HST Cepheid recalibrated TF scale at v$\sim$1000 kms$^{-1}$
certainly brackets values as low as H$_0\sim$40 kms$^{-1}$Mpc$^{-1}$. Thus we
take the view that the Tully-Fisher estimate of H$_0$ may not yet have converged to its
final value and that the Tully-Fisher relation may still allow our simple cosmological
model with $\Omega_{baryon}$=1.


\end{document}